\documentclass[prmaterials,aps,amsmath,amssymb,twocolumn,showpacs,superscriptaddress]{revtex4-2}

\usepackage{graphicx}
\usepackage[colorlinks=true, citecolor=blue, urlcolor=blue, linkcolor=blue, hypertexnames=false]{hyperref}
\usepackage{placeins}
\makeatletter
\def\fnum@figure{\textbf{\figurename\nobreakspace\thefigure}}
\makeatother

\makeatletter
\def\doauthor#1#2#3{%
  \ignorespaces#1\unskip
  \begingroup
   #3%
  \@if@empty{#2}{\endgroup{}{}}{\endgroup{\comma@space}{}\frontmatter@footnote{#2}}%
  \@listcomma
  \space \@listand
}%
\makeatother

\begin{document}

\title{Intrinsically low thermal conductivity of stoichiometric lithium niobate: \\ Experimental measurement and microscopic origin}

\author{Wenjiang Zhou}
\thanks{These authors contributed equally to this work.}
\affiliation{School of Mechanics and Engineering Science, Peking University,
  Beijing 100871, China}

\author{Fuwei Yang}
\thanks{These authors contributed equally to this work.}
\affiliation{National Key Laboratory of Advanced Micro and Nano Manufacture
  Technology, Peking University, Beijing 100871, China}
\affiliation{Department of Engineering Mechanics, Tsinghua University, Beijing
  100084, China}

\author{Yuxi Wang}
\affiliation{School of Mechanics and Engineering Science, Peking University,
  Beijing 100871, China}

\author{Weiheng Li}
\affiliation{School of Mechanics and Engineering Science, Peking University,
  Beijing 100871, China}

\author{Wujuan Yan}
\affiliation{School of Mechanics and Engineering Science, Peking University,
  Beijing 100871, China}

\author{Kexin Zhang}
\affiliation{School of Physics, Peking University, Beijing 100871, China}

\author{Bai Song}
\email{songbai@pku.edu.cn}
\affiliation{School of Mechanics and Engineering Science, Peking University,
  Beijing 100871, China}
\affiliation{National Key Laboratory of Advanced Micro and Nano Manufacture
  Technology, Peking University, Beijing 100871, China}

\date{\today}

\begin{abstract}
With the rapid development of integrated electro-optic and nonlinear optical devices based on lithium niobate (LiNbO$_3$, LN), thermal management is becoming a critical area of focus. However, experimental measurement of thermal transport in stoichiometric LiNbO$_3$ (sLN) remains scarce, and the intrinsic microscopic mechanisms remain to be established. Here, we combine the laser pump-probe technique of frequency-domain thermoreflectance (FDTR) with state-of-the-art machine-learned atomistic simulations to comprehensively investigate thermal transport in sLN. The measured and simulated room-temperature thermal conductivity ($\kappa$) values of sLN agree well, which are orders-of-magnitude lower than that of many classic and emerging semiconductors such as silicon. Furthermore, the temperature-dependent $\kappa$ exhibits a $T^{-\alpha}$ scaling with $\alpha$ near unity, suggesting that thermal transport is dominated by intrinsic phonon-phonon scattering. By comparing sLN with cubic boron arsenide (cBAs) which serves as an ultrahigh-$\kappa$ benchmark, we reveal that harmonic properties are not responsible for the low $\kappa$ of sLN, which feature phonon heat capacity and group velocities that are either higher than or comparable to those in cBAs. Instead, the low $\kappa$ originates from substantially stronger anharmonicity and larger scattering phase space. These two factors collectively suppress phonon lifetimes by 1--2 orders of magnitude, leading to a maximum phonon mean free path of approximately 140~nm. As a result, notable size effects emerge in thin-film sLN below 1~$\mu$m, with $\kappa$ dropping to half the bulk value at 10~nm. Altogether, our findings establish a fundamental understanding of thermal transport in sLN and provide atomistic insights for thermal management in advanced lithium niobate technologies.
\end{abstract}

\maketitle

\section{Introduction}
\begin{figure}
\centering
\includegraphics[width=0.85\columnwidth]{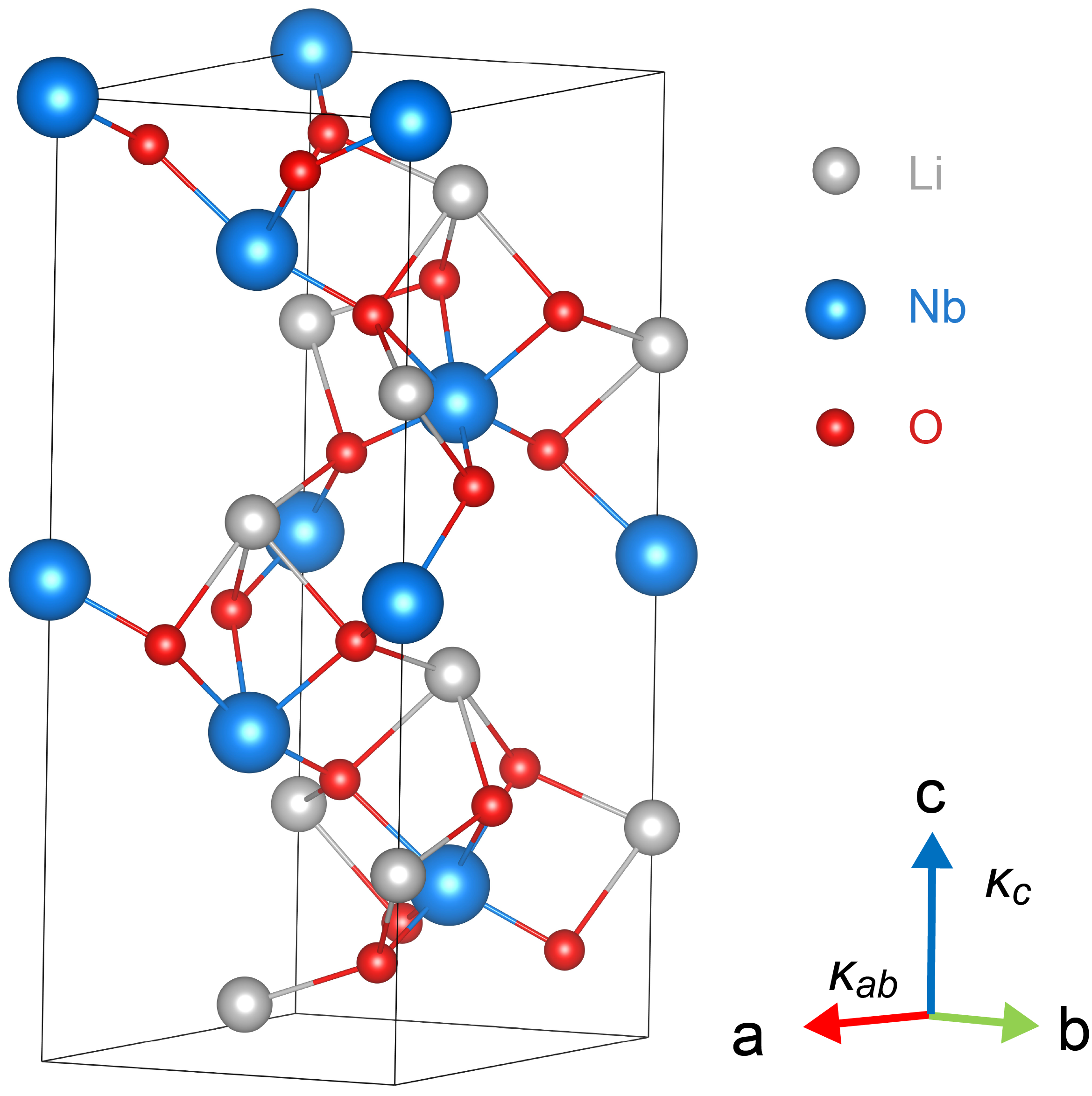}
\caption{\textbf{Atomic structure of sLN.} The conventional cell is shown, with Li,
  Nb, and O atoms occupying distinct crystallographic sites in the $R3c$ space
  group. The in-plane thermal conductivity is isotropic, denoted as $\kappa_{ab}$, and the out-of-plane component is $\kappa_{c}$.}
\label{fig:structure}
\end{figure}

Since its discovery in 1949~\cite{Matthias1949}, lithium niobate (LiNbO$_3$, LN) has been established as an extremely versatile functional oxide with large electro-optic, piezoelectric, and nonlinear optical coefficients, a wide transparency window, long-term stability~\cite{Weis1985}, and the ability to be periodically poled~\cite{myers1995quasi,hu2025integrated}. Together with the commercial availability and low fabrication cost of wafer-scale crystals, these characteristics have supported the broad use of LN in electro-optic
modulators~\cite{Wang2018Integrated,zhang2019broadband, Li2020}, nonlinear
wavelength conversion devices~\cite{Wang2018Optica,lin2019broadband},
integrated photonic platforms~\cite{zhu2021integrated,boes2023lithium,feng2024integrated,stokowski2024integrated},
quantum light sources~\cite{zhu2025integrated}, surface acoustic wave filters~\cite{zheng2023near}, and sub-ångström snapshot spectroscopy~\cite{yao2025integrated}. In particular, thin-film LN has enabled nanoscale optical confinement and enhanced electro-optic interaction while maintaining low optical losses~\cite{levy1998fabrication,rabiei2004optical,zhu2021integrated, chen2022advances,boes2023lithium}.
Within the LN material family, congruent LiNbO$_3$ (cLN) with a Li-to-Nb atomic ratio of 48.6/51.4 remains the most widely explored member, although it contains intrinsic nonstoichiometric point defects such as Li-site
vacancies and antisite Nb atoms. With a near-ideal Li-to-Nb ratio and reduced point defects, stoichiometric LiNbO$_3$ (sLN, Fig.~\ref{fig:structure}) has recently attracted growing attention due to the larger electro-optic and nonlinear optical coefficients together with a high Curie temperature, making it a superior material platform for high-performance devices \cite{fujiwara1999comparison, klein2003absolute, Nakamura2008}.

With the rapid development LN devices toward higher levels of integration and higher-frequency operation, thermal management is becoming a critical area of focus. To date, experimental studies of thermal transport in LN have focused mainly on cLN, reporting room-temperature thermal conductivity ($\kappa$) values around 2--4~Wm$^{-1}$K$^{-1}$~\cite{Yao2008,ziade2020wide,Bashir2024,bao2025machine}. In contrast, experimental results on the thermal conductivity of sLN remain scarce, despite its great potential. To date, the prevailing benchmark remains a laser-flash measurement performed nearly two decades ago, which reported $\kappa_{ab}=4.4$~Wm$^{-1}$K$^{-1}$ and $\kappa_{c}=5.2$~Wm$^{-1}$K$^{-1}$ at room temperature~\cite{Yao2008}. However, this method reflects the thermal conductivity across a macroscopic length scale and may therefore be influenced by extended structural imperfections and sample inhomogeneity. Consequently, the intrinsic thermal limits of sLN remain poorly defined, a gap that can only be resolved through spatially resolved thermal characterizations and simulations at the micro- and atomic scales.

Theoretically, atomistic simulations have provided useful insights into thermal transport in LN. First-principles calculations combined with the phonon Boltzmann transport equation (BTE) considering only three-phonon interactions showed that the thermal conductivity anisotropy originates from different group velocities along the $ab$--plane and $c$--direction, and also highlighted the non-negligible contribution of optical phonon branches to $\kappa$~\cite{Fu2021}. More recently, the effects of four-phonon scattering and off-diagonal wave-like contribution have also been investigated~\cite{Ju2025}. Nevertheless, the microscopic mechanisms of thermal transport in sLN has not been fully understood. In particular, it is unclear whether the low $\kappa$ is mainly limited by harmonic properties or by anharmonic effects. Furthermore, because thin-film LN devices typically operate at nanoscale dimensions (tens to hundreds of nanometers), mapping the phonon mean free path spectrum and size-dependent thermal conductivity is imperative for device design. Despite this necessity, quantitative data remain surprisingly scarce.

In this work, we first measure the temperature-dependent thermal conductivity of single-crystal sLN using frequency-domain thermoreflectance (FDTR), a non-contact optical method that is capable of probing thermal transport at the micro- and nanoscale. Then, we perform state-of-the-art machine-learned molecular dynamics simulations to calculate the thermal conductivity. The computational results agree well with our experimental measurements, confirming the intrinsically low $\kappa$ of sLN. Furthermore, to understand the origin of the low $\kappa$, we perform systematic harmonic and anharmonic analyses, by comparing the heat capacity, group velocity, scattering phase space, anharmonic factor, and phonon lifetime of sLN with those of cubic boron arsenide (cBAs). We choose cBAs as a benchmark because it is a prototypical material with an ultrahigh lattice $\kappa$ of $\sim$1300~Wm$^{-1}$K$^{-1}$~\cite{tian2018unusual,li2018high,kang2018experimental}. Our results reveal that while the harmonic properties of sLN are favorable for high $\kappa$, the enhanced scattering phase space and strong anharmonicity substantially suppress the phonon lifetime and thus thermal transport. Finally, we employ non-equilibrium molecular dynamics (NEMD) and homogeneous non-equilibrium molecular dynamics (HNEMD) simulations to investigate the thermal conductivity of thin-film sLN with different thicknesses.

\section{Methods}

\subsection{Thermal transport measurement}

\begin{figure}[ht]
\centering
\includegraphics[width=\columnwidth]{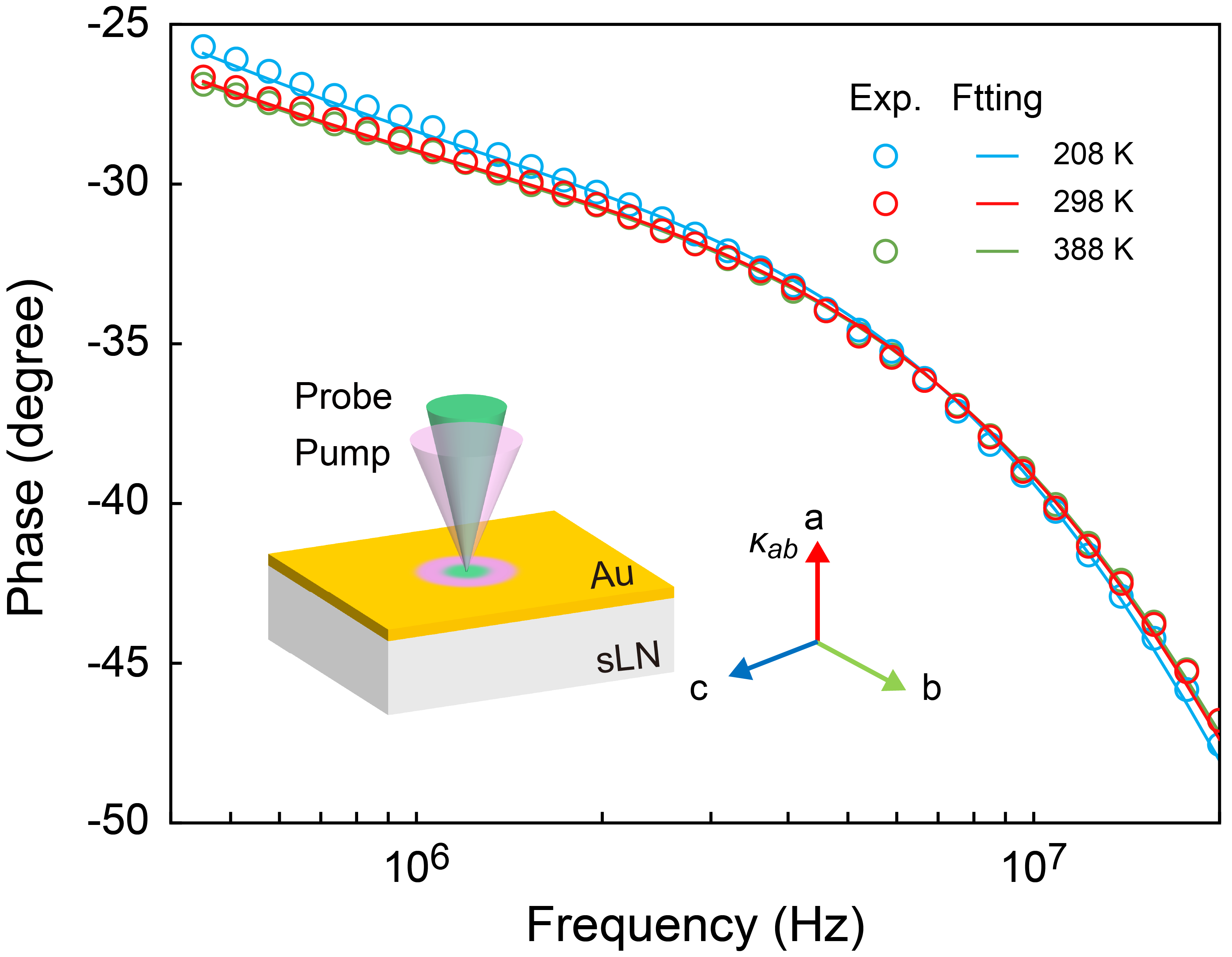}
\caption{\textbf{Thermal transport measurement in sLN.} Raw and fitted thermoreflectance phase signals of sLN as a function of modulation frequency. Inset: Schematic of the
  FDTR measurement configuration wth respect to the crystallographic axes of sLN. Since the sample is X-cut, the
  out-of-plane transport direction corresponds to the $a$-- or $b$--axis.}
\label{fig:fdtr}
\end{figure}

Our FDTR measurements were performed on sLN single crystals obtained from Jiangxi Unicrystal Technology. The sample was obtained from a 2-inch X-cut wafer and was oriented with the $ab$ crystallographic plane parallel to the out-of-plane direction in our experimental setup (inset of Fig.~\ref{fig:fdtr}). We coated the sample surface with a gold (Au) transducer layer via electron beam evaporation, which serves as the optical absorber and heat source in the FDTR measurement. Two continuous-wave lasers were employed in our platform as the pump and probe. The 405 nm pump laser was modulated at frequency $f$ ranging from 400~kHz to 20~MHz, inducing a periodic temperature variation in the sample. The resulting temperature oscillations were monitored by the probe laser (532~nm) through the thermoreflectance of the Au transducer, whose phase and amplitude as a function of $f$ were resolved by a lock-in amplifier. The effective laser spot size was determined to be 2.9~$\mu$m. The sample temperature was controlled via a V-200 cryostat from Physike. At each temperature, the FDTR signals from 10 different spots were acquired and averaged. More details about the platform can be found in our previous works~\cite{Yang2025PRL,Yang2025PRB,RN1192,yan2025thermal}. 

To extract $\kappa$ of sLN, the measured phase data were fitted to a multi-layer anisotropic heat conduction model~\cite{cahill2004analysis,schmidt2009frequency}. Among the key input parameters, the thermal conductivity of the Au transducer was independently determined by combining four-probe electrical resistivity measurement with the Wiedemann-Franz law, yielding $\kappa_{\rm Au} = 180$~Wm$^{-1}$K$^{-1}$ at room temperature. Measurements at other temperatures were also performed. The Au thickness was measured by atomic force microscopy (AFM) on a reference sample coated together with the sLN sample. The heat capacity of sLN was obtained from first-principles calculations with full temperature dependence. Sensitivity analysis shows that the phase response has negligible sensitivity to $\kappa_c$ over the measured frequency range (Fig.~\ref{fig:fdtr_sen} in the Supplemental Material~\cite{si}). Therefore, $\kappa_c$ was fixed during the fitting, and only $\kappa_{ab}$, together with the Au/sLN interfacial thermal conductance $G$, were treated as the unknowns to be fitted.

\subsection{Computational methods}

\subsubsection{Dataset and training for machine-learned potentials}

To enable large-scale molecular dynamics (MD) simulations of high accuracy, we first constructed a training dataset for the neuroevolution potential (NEP)~\cite{fan2021neuroevolution} from \textit{ab initio} molecular dynamics (AIMD) calculations covering a temperature range of 100--1200~K. This was supplemented by perturbed configurations including random atomic displacements and cell deformations with tensile and compressive strains up to $\pm3\%$. Each structure was generated using $2\times2\times2$ supercells of the rhombohedral primitive cell, containing 80 atoms. The resulting training and test set contains 615 and 153 structures, respectively.

Density functional theory (DFT) calculations for the reference data were performed using the Vienna \textit{ab initio} simulation package (VASP)~\cite{kresse1996efficient} with the projector-augmented wave (PAW)
  method~\cite{blochl1994projector}. The Perdew-Burke-Ernzerhof (PBE) generalized gradient approximation~\cite{perdew1996generalized} was employed for the exchange-correlation functional, with a plane-wave energy cutoff of 600~eV and a $\Gamma$-centered $k$-point mesh of $3\times3\times3$. The energy convergence criterion was set to $10^{-8}$~eV.

Based on this dataset, a NEP model~\cite{fan2021neuroevolution,dong2024molecular,xu2025gpumd} was trained using the GPUMD package~\cite{fan2017efficient}. The cutoff radii for both the radial and angular descriptor components were set to 6~\AA, and the network employed a single hidden layer with 50 neurons. The training was performed using the separable natural evolution strategy~\cite{Schaul2011} to minimize a loss function that simultaneously considers the root-mean-square errors (RMSEs) of energy, force, and virial stress, with $L_1$ and $L_2$ regularization. The resulting RMSEs on the test set are 0.65~meV/atom, 51.65~meV/\AA{}, and 6.83~meV/atom for energy, force, and virial, respectively, with the corresponding parity plots (Fig.~\ref{fig:parity}) shown in the Supplemental Material~\cite{si}.

To assess the reliability of the PBE functional for describing the crystal structure of sLN, we compared the fully relaxed lattice constants with experimentally measured values~\cite{RN1265}. The experimental lattice constants are $a=5.15$~\AA{} and $c=13.86$~\AA{}, while our PBE-optimized values are $a=5.18$~\AA{} and $c=13.99$~\AA{}, corresponding to deviations of approximately 0.6\% and 0.9\%, respectively. For further validation, we performed path-integral molecular dynamics (PIMD) simulations~\cite{ying2025highly} at 300~K using the NEP model, which yielded lattice constants of $a=5.24$~\AA{} and $c=14.03$~\AA{}, accounting for thermal expansion and nuclear quantum effects. The small deviations from the experimental values indicate that both the PBE functional and the trained NEP model can accurately capture the structural properties of sLN.

To further validate the trained potential, we calculated the phonon dispersion of sLN using the temperature dependent effective potential (TDEP) method~\cite{Hellman2011,Knoop2024} based on both DFT and the NEP model using the primitive cell. In the phonon calculations, we employed density functional perturbation theory (DFPT)~\cite{Baroni2001} to calculate the Born effective charges and high-frequency dielectric constant, which were used to account for the long-range Coulomb interactions and the longitudinal and transverse optical (LO-TO) phonon splitting~\cite{Gonze1994, Gonze1997}. As shown in Fig.~\ref{fig:phonon}, the phonon dispersions obtained from DFT and NEP are in excellent agreement with each other, and the optical phonon frequencies at the Brillouin zone center are consistent with Raman measurements~\cite{Claus1972, Ridah1997}.

\subsubsection{HNEMD simulations}

Although the BTE approach can be employed to compute and analyze the lattice thermal conductivity of sLN, our calculations primarily rely on ML-assisted HNEMD, which non-perturbatively captures phonon scattering to all orders and offers high robustness across different system configurations~\cite{zhou2025heat}. In the HNEMD approach, an external driving force characterized by the parameter $\mathbf{F}_{\rm e}$ is applied to each atom as~\cite{Wu2024JCP}:
\begin{equation}
\mathbf{F}_i^{\rm ext} = \mathbf{F}_{\rm e} \cdot \mathbf{W}_i.
\end{equation}
Here, $\mathbf{W}_i$ is the virial tensor of atom $i$ defined as~\cite{Wu2024JCP}:
\begin{equation}
\mathbf{W}_i = \sum_{j\neq i} \mathbf{r}_{ij} \otimes \frac{\partial
  U_j}{\partial \mathbf{r}_{ji}},
\end{equation}
where $U_j$ is the site potential of atom $j$, $\mathbf{r}_i$ and $\mathbf{r}_j$ are the positions of atoms $i$ and $j$, respectively, and $\mathbf{r}_{ij} \equiv
  \mathbf{r}_j - \mathbf{r}_i$. The thermal conductivity tensor $\kappa_{\mu\nu}$ can be computed from:
\begin{equation}
\frac{\langle J_\mu(t) \rangle_{\rm ne}}{TV} = \sum_\nu \kappa_{\mu\nu} F_{\rm
  e}^\nu,
\end{equation}
where $\langle J_\mu(t) \rangle_{\rm ne}$ is the non-equilibrium ensemble average of the heat current, $T$ is the absolute temperature, and $V$ is the system volume. The heat current is given by $\mathbf{J} =
  \sum_i \mathbf{W}_i \cdot \mathbf{v}_i$, where $\mathbf{v}_i$ is the velocity of atom $i$.

Furthermore, the HNEMD formalism yields the frequency-resolved thermal conductivity $\kappa_{\mu\nu}(\omega)$ via the virial-velocity correlation function:
\begin{equation}
K_\mu(t) = \sum_i \sum_\nu \langle W_i^{\mu\nu}(0) v_i^\nu(t) \rangle_{\rm ne},
\label{eq:Kt}
\end{equation}
which leads to the spectral decomposition relation:
\begin{equation}
\frac{2}{VT} \int_{-\infty}^{+\infty} e^{i\omega t} K_\mu(t) \, dt =
\sum_\nu \kappa_{\mu\nu}(\omega) F_{\rm e}^\nu.
\end{equation}

The HNEMD simulations were performed on supercells containing 15,000 atoms. The time step was set to 1~fs. To ensure reliable results, the driving force parameter was carefully tested for each temperature to achieve a high signal-to-noise ratio while maintaining the system in the linear-response regime (Fig.~\ref{fig:hnemd_fe} in the Supplemental Material~\cite{si}). Each system was first equilibrated in the $NPT$   ensemble~\cite{Berendsen1984} at zero pressure for 0.5~ns and then in the $NVT$ ensemble for another 0.5~ns using the Nos\'e-Hoover chain thermostat~\cite{Martyna1992}, prior to the production runs of 10~ns in the $NVT$ ensemble. Error bars were estimated from the standard deviation of five independent simulations in the second half of the production run. Unless otherwise specified, isotope scattering was not considered in all MD simulations~\cite{zhou2024isotope,zhou2024prb}. For comparison, we adopted the results for cBAs from our previous work~\cite{zhou2025insight}, which were obtained using lattice dynamics simulations.

\subsubsection{NEMD simulations}
To investigate the size-dependent thermal transport in thin-film sLN, we performed ML-assisted NEMD
  simulations~\cite{dong2024molecular,zhou2025ultrahigh}. In this approach, a pair of heat source and sink was implemented using Langevin thermostats~\cite{Bussi2007}, with fixed boundary conditions applied along the transport direction and periodic boundary conditions in the transverse directions (Fig.~\ref{fig:nemd} in the Supplemental Material~\cite{si}). Upon reaching steady state, a temperature profile with a gradient $\nabla T$ is established, and a steady heat flux $Q$ is generated. The apparent thermal conductivity of a finite system is then obtained from Fourier's law:
\begin{equation}
\kappa(L) = \frac{Q}{|\nabla T|}.
\end{equation}
The heat flux is computed from the energy exchange rate in the thermostatted regions as $Q = (1/A)\, dE/dt$, where $A$ is the cross-sectional area.

The NEMD formalism also enables spectral decomposition of the thermal conductance via the virial-velocity correlation function $K_\mu(t)$ [defined in Eq.~\ref{eq:Kt}], which is now evaluated under the NEMD simulations. The frequency-resolved thermal conductance $G(\omega)$ is then given by:
\begin{equation}
G(\omega) = \frac{2}{V \Delta T} \int_{-\infty}^{+\infty} e^{i\omega t}
K_\mu(t) \, dt,
\end{equation}
which satisfies $G = \int_{-\infty}^{+\infty} G(\omega) \, d\omega / 2\pi$.

We performed NEMD simulations for four system lengths ranging from 1.3~nm to 46.5~nm. Each system was first relaxed in the $NPT$ ensemble~\cite{Berendsen1984} at 300~K for 0.1~ns and then equilibrated in the $NVT$ ensemble for another 0.1~ns, using a time step of 1~fs. A temperature difference of $\pm10$~K was then applied via Langevin thermostats~\cite{Bussi2007} at the heat source and sink, and production runs of 2~ns were carried out to collect the steady-state heat flux and temperature profile.

\subsubsection{Spectral energy density simulations}

 Spectral energy density (SED) analysis~\cite{Thomas2010,Liang2025} was employed to extract the phonon lifetimes. The SED approach was based on the conventional cell of sLN, while MD trajectory sampling was performed on $20\times20\times10$ and $6\times6\times100$ supercells for modes with atomic vibrations within the $ab$--plane and along the $c$-direction, respectively. The time step was also chosen as 1~fs. Each system was first relaxed in the $NPT$ ensemble at 300~K and zero pressure for 0.2~ns, then equilibrated in the $NVT$ ensemble for another 0.2~ns. The production run was then carried out in the $NVE$ ensemble, during which atomic velocities are collected every 15~fs, yielding a total of 20,000 snapshots. These velocities were projected onto phonon mode coordinates, and the SED results were fitted to Lorentzian functions to extract phonon frequencies and lifetimes.

\section{Results and Discussion}

\subsection{Intrinsically low thermal conductivity of sLN}

\begin{figure*}
\centering
\includegraphics[width=0.95\textwidth]{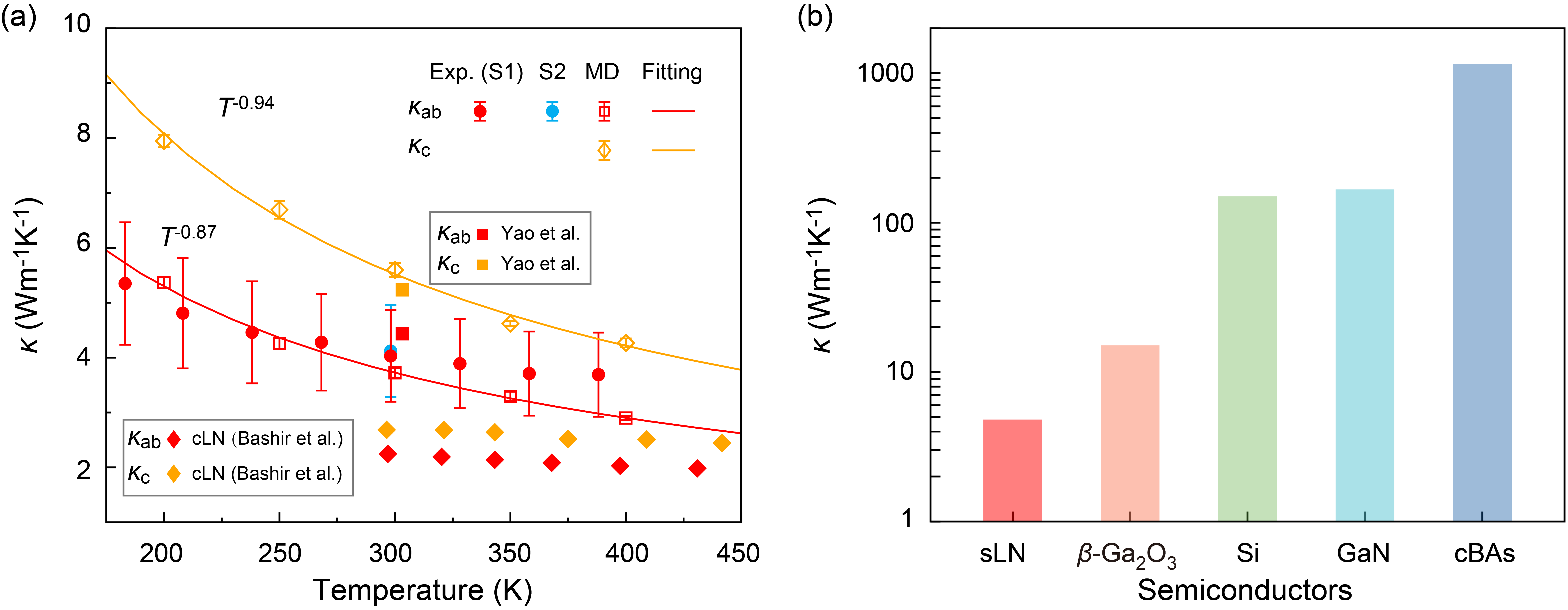}
\caption{\textbf{Thermal conductivity of sLN.} (a) FDTR measured $\kappa_{ab}$ and HNEMD calculated $\kappa_{ab}/\kappa_{c}$ of sLN as a function of temperature. For comparison, literature data are plotted including temperature-dependent measurements on cLN by Bashir~\textit{et~al.}~\cite{Bashir2024} and two data points on
  sLN measured at room-temperature by Yao~\textit{et~al.}~\cite{Yao2008}. Our single-crystal sLN sample, denoted S1, was measured from about 183 to 388 K, while sample S2 was only used for validation at room temperature. (b) Room-temperature $\kappa$ of sLN compared with classic and emerging semiconductor materials. Data for cBAs, $\beta$-Ga$_2$O$_3$, GaN, and Si are taken from Refs.~\cite{tian2018unusual,li2018high,kang2018experimental,Jiang2018,RN1192,Shanks1963}. The $\kappa$ values shown for sLN and $\beta$-Ga$_2$O$_3$ are the orientation-averaged thermal conductivities.}
\label{fig:kappa}
\end{figure*}

In Fig.~\ref{fig:fdtr}, we present the FDTR phase signals measured on sLN sample S1 at three representative temperatures (208~K, 298~K, and 388~K), together with the corresponding fitted curves. The fitted and measured phase responses agree well over the full frequency range. The room-temperature thermal conductivity in the $ab$--plane is extracted to be $\kappa_{ab}=4.0\pm0.8$~W~m$^{-1}$~K$^{-1}$. A second sample (S2) gives a value of $\kappa_{ab}=4.1\pm0.8$~W~m$^{-1}$~K$^{-1}$, indicating good sample-to-sample consistency and reproducibility. These FDTR results are close to the previously reported value of $\kappa_{ab}=4.4$~W~m$^{-1}$~K$^{-1}$ for near-stoichiometric LN measured by the laser-flash method~\cite{Yao2008}. The agreement between the local FDTR measurement and the earlier macroscopic laser-flash measurement suggests that the low thermal conductivity is not limited by the sample quality. This conclusion is further supported by our ML-assisted HNEMD calculations. At 300~K, the calculated in-plane thermal conductivity is $\kappa_{ab}=3.7\pm0.1$~W~m$^{-1}$~K$^{-1}$, in close agreement with the FDTR measurements. In addition, the calculated $\kappa_c$ is $5.6\pm0.1$~W~m$^{-1}$~K$^{-1}$, also consistent with the value of $5.2$~W~m$^{-1}$~K$^{-1}$ reported by Yao~\textit{et~al.}~\cite{Yao2008}. The combined experimental and computational results therefore demonstrate that the low thermal conductivity of sLN is an intrinsic material property.

In addition to room-temperature investigation, we further characterized and calculated the temperature dependence of thermal transport in sLN. As shown in Fig.~\ref{fig:kappa}(a), the measured $\kappa_{ab}$ agrees well with the simulation, though it is slightly smaller at low temperatures and larger at high temperatures. This modest discrepancy likely reflects experimental uncertainties and the neglect of nuclear quantum effects in the simulation. The thermal conductivity of sLN exhibits a monotonic decreasing trend with increasing temperature. We fitted the simulation results to $\kappa\propto T^{-\alpha}$, which yields an exponent close to $-1$, with $\alpha=0.87$ for $\kappa_{ab}$ and $\alpha=0.94$ for $\kappa_{c}$ between 200 and 450~K, suggesting that intrinsic anharmonic phonon-phonon scattering dominates thermal transport in sLN. To make a comparison, we also included the measured data of cLN~\cite{Bashir2024} in Fig.~\ref{fig:kappa}(a). The $\kappa$ of sLN is consistently higher than that of cLN across the measured range, which reflects the reduced point-defect scattering in the stoichiometric composition~\cite{bao2025machine}.

With the technological relevance of sLN in mind, we then place the $\kappa$ of sLN in the broader context of several classic and emerging semiconductor materials including gallium oxide ($\beta$-Ga$_2$O$_3$), silicon (Si), gallium nitride (GaN), and cBAs. As shown in Fig.~\ref{fig:kappa}(b), sLN exhibits the lowest thermal conductivity, with its room-temperature $\kappa$ approximately one order of magnitude lower than that of $\beta$-Ga$_2$O$_3$, an ultra-wide-bandgap semiconductor already known for its characteristically low $\kappa$. This comparison motivates a microscopic analysis into the origin of the intrinsically low thermal conductivity of sLN. In the following, we address this question through a comparative study between sLN and cBAs.

\subsection{Phonon heat capacity and group velocity}

\begin{figure*}
\centering
\includegraphics[width=0.95\textwidth]{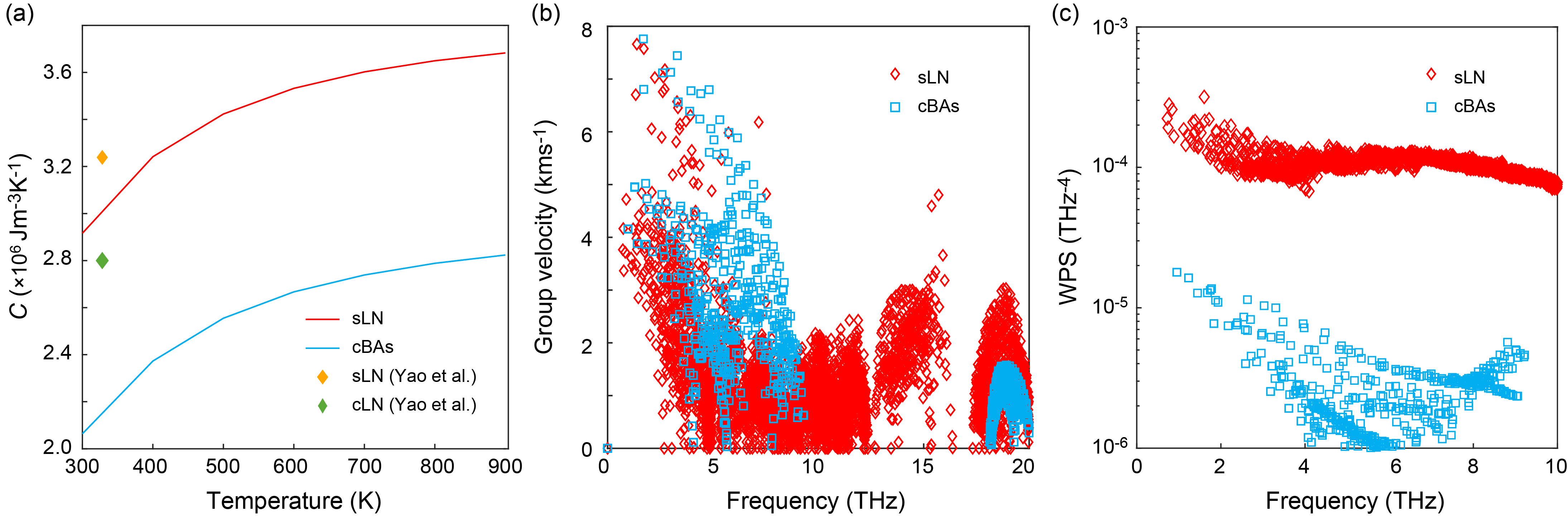}
\caption{\textbf{Comparison of heat capacity, group velocity, and phonon scattering phase space between sLN and cBAs.} (a) Volumetric heat capacity
  as a function of temperature. The measured data for sLN are taken from Yao~\textit{et~al.}~\cite{Yao2008}. (b) Phonon group velocities as a function of
  frequency. (c) Phonon scattering phase space as a function of frequency,
  showing substantially enhanced scattering channels in sLN.}
\label{fig:harmonic}
\end{figure*}

The lattice thermal conductivity can be expressed as $\kappa \propto \sum_{\mathbf{q}\nu} C_{\mathbf{q}\nu} v_{\mathbf{q}\nu}^2
  \tau_{\mathbf{q}\nu}$, where $C_{\mathbf{q}\nu}$ is the volumetric heat capacity of phonon mode $\nu$ at wave vector $\mathbf{q}$, $v_{\mathbf{q}\nu}$ is the phonon group velocity, and $\tau_{\mathbf{q}\nu}$ is the lifetime. The heat capacity and group velocity depend only on the harmonic (second-order) interatomic force constants, whereas the lifetime is governed by the phonon scattering phase space and the anharmonicity. To determine which factor limits thermal transport in sLN, we first compare its harmonic properties with those of cBAs~\cite{tian2018unusual,li2018high,kang2018experimental}, a prototypical ultrahigh-$\kappa$ material with a room-temperature $\kappa$ around $1300$~Wm$^{-1}$K$^{-1}$, exceeding that of sLN by over two orders of magnitude.

The mode-resolved phonon heat capacity and group velocity were calculated using the ShengBTE package~\cite{Li2014} with the harmonic (second-order) interatomic force constants obtained from the TDEP method~\cite{Hellman2011,Knoop2024}. For cBAs, the results are taken from our previous work~\cite{zhou2025insight}. Interestingly, as shown in Fig.~\ref{fig:harmonic}(a), the total heat capacity $C = \sum_{\mathbf{q}\nu} C_{\mathbf{q}\nu}$ of sLN is approximately 40\% higher than that of cBAs across the entire temperature range. Further, we plot for both sLN and cBAs the group velocities of all phonon branches as a function of frequency in Fig.~\ref{fig:harmonic}(b). The group velocities of cBAs are slightly higher than those of sLN, and the maximum values of both materials reach approximately 8~km/s for the acoustic phonon branches. These results indicate that heat capacity and group velocity are not the limiting factors for the low $\kappa$ of sLN.

\subsection{Anharmonicity-suppressed thermal transport}

\begin{figure*}
\centering
\includegraphics[width=0.95\textwidth]{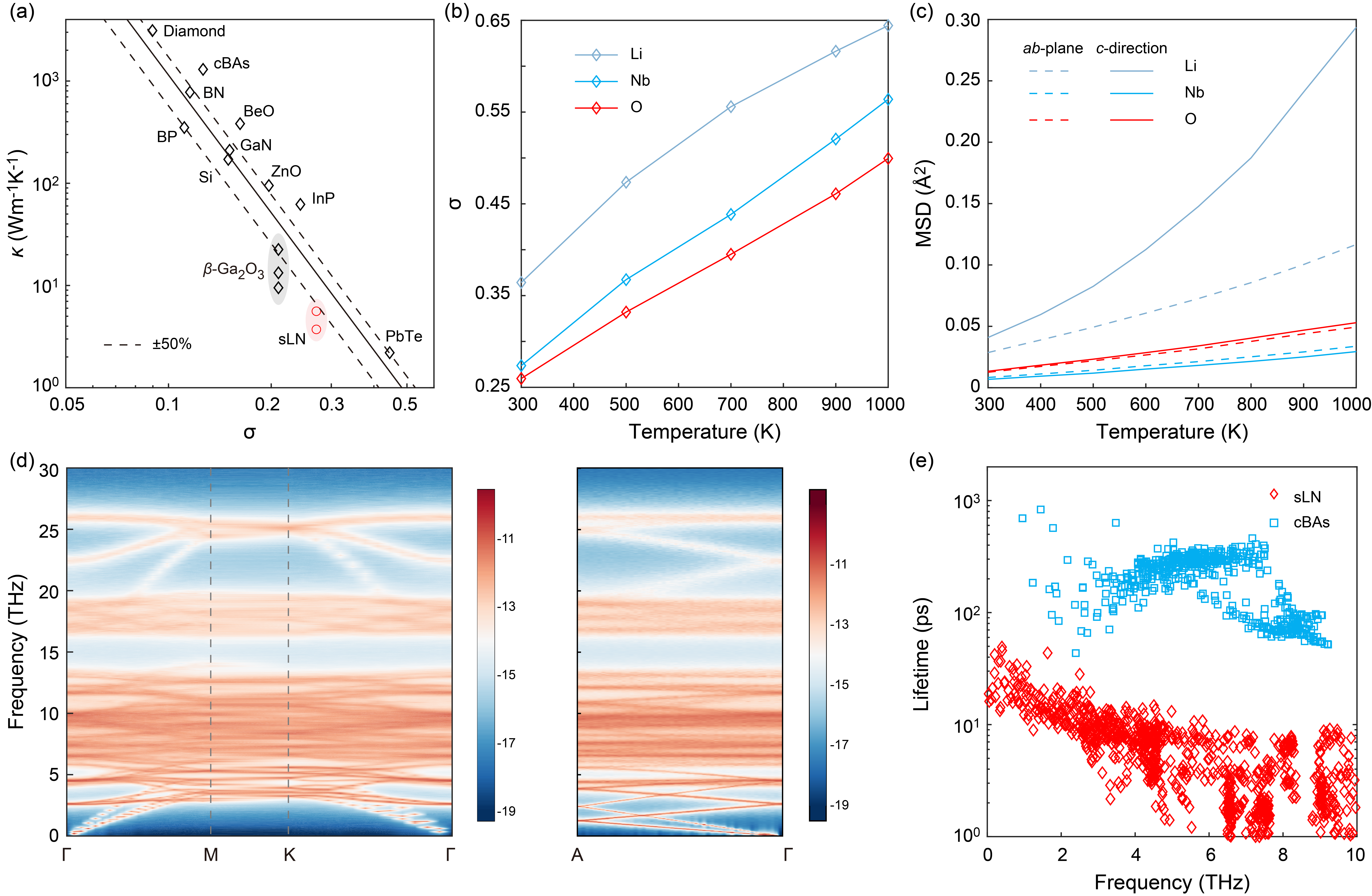}
\caption{\textbf{Anharmonic analysis of sLN.} (a) Thermal conductivity of representative
  materials as a function of the anharmonic factor $\sigma_{\mathrm{A}}$. The $\sigma_{\mathrm{A}}$ data for cBAs and sLN are from this work, while those for other materials are taken from Refs.~\cite{Knoop2023,Zeng2024,Alkandari2025}. (b)
  Atom-resolved anharmonic factor $\sigma_{\mathrm{A}}$ as a function of
  temperature. (c) Mean squared displacement (MSD) as a function of temperature.
  (d) Spectral energy density (SED) for phonon modes within the $ab$--plane and along the $c$-direction, computed on the conventional cell. (e) Extracted phonon lifetimes $\tau_{\mathbf{q}\nu}$ as functions of frequency, compared with cBAs.}
\label{fig:anharmonic}
\end{figure*}

Beyond heat capacity and group velocity, phonon lifetime fundamentally dictates thermal conductivity. The lifetime is governed by two factors: The phonon scattering phase space, which quantifies the number of available scattering channels that simultaneously satisfy energy and crystal-momentum conservation; and the anharmonicity of the interatomic forces, which determines the magnitude of the phonon scattering coupling matrix elements. As an example, we evaluate the three-phonon scattering phase space according to the formulation in Ref.~\cite{lindsay2008three}. As plotted in Fig.~\ref{fig:harmonic}(c), sLN exhibits a substantially larger phonon scattering phase space than cBAs across all phonon frequencies. This dramatically enlarged scattering phase space originates from two structural features: (i) The larger number of atoms in the primitive cell of sLN (10 versus 2 in cBAs), which generates a much higher
density of optical phonon branches that provide abundant scattering channels; and (ii) the lower lattice symmetry (rhombohedral $R3c$ space group) of sLN, which relaxes the symmetry-based selection rules that restrict scattering processes in cBAs. These findings are consistent with recent studies on the role of crystal symmetry in thermal transport of materials such as rhombohedral boron nitride~\cite{Yang2025PRB} and wurtzite GaN~\cite{quan2021electric}.

To quantify the degree of anharmonicity, Knoop~\textit{et~al.}~\cite{Knoop2020} originally introduced the anharmonic factor $\sigma_{\mathrm{A}}$ based on zero-temperature force constants:
\begin{equation}
\sigma_{\mathrm{A}} = \sqrt{\frac{\sum_{i\alpha} \left\langle \left[
  F_{i\alpha} - F_{i\alpha}^{(2)} \right]^{2} \right\rangle}{\sum_{i\alpha}
  \left\langle (F_{i\alpha})^{2} \right\rangle}},
\end{equation}
where $F_{i\alpha}$ denotes the force component on atom $i$ in direction $\alpha$, $F_{i\alpha}^{(2)}$ is the second-order (harmonic) force component, and $\langle\cdots\rangle$ denotes the ensemble average. This quantity can be further decomposed into contributions from the Li, Nb, and O atoms. Building on this concept, Alkandari~\textit{et~al.}~\cite{Alkandari2025} later proposed a temperature-dependent variant. Similarly, we adopt the temperature-dependent effective potential method~\cite{Hellman2011,Knoop2024} to compute $F_{i\alpha}^{(2)}$ at finite temperatures. 

We first examine the relation between anharmonicity and thermal conductivity. As demonstrated in Fig.~\ref{fig:anharmonic}(a) for materials of very low to ultrahigh thermal conductivity, $\kappa$ decreases monotonically with increasing $\sigma_{\mathrm{A}}$, suggesting that stronger anharmonicity systematically suppresses thermal transport. The $\sigma_{\mathrm{A}}$ of sLN at 300~K is approximately 0.27, which is substantially stronger than that of cBAs ($\sigma_{\mathrm{A}}\approx0.12$).  We further consider the temperature evolution of $\sigma_{\mathrm{A}}$ [Fig.~\ref{fig:anharmonic}(b)]. Over the range of 300--1000~K, the atomically-resolved $\sigma_{\mathrm{A}}$ consistently increases with temperature, as expected for thermally activated anharmonic phonon-phonon interactions. Comparing different atomic species, Li atoms possess the largest anharmonic factor, approximately 40\% larger than those of Nb and O atoms across the entire temperature range. Consistently, Li atoms also exhibit the largest mean squared displacement (MSD) [Fig.~\ref{fig:anharmonic}(c)], reflecting their dominant role in anharmonic phonon scattering. This is likely attributable to the light atomic weight of Li and its loosely bound position within the oxygen octahedral framework.

Having analyzed the harmonic properties and anharmonicity separately, we now examine their combined effect on the mode-resolved phonon lifetimes $\tau_{\mathbf{q}\nu}$. The SED simulations [Fig.~\ref{fig:anharmonic}(d)], as described in the Methods, were performed to extract phonon lifetimes [Fig.~\ref{fig:anharmonic}(e)]. We focus on the 0--10~THz range, as these low-frequency phonons are the dominant heat carriers. The SED spectra exhibit well-defined phonon features at room temperature, which indicates that the particle assumption for phonon transport is valid in sLN. Owing to the combined effects of the enlarged scattering phase space and stronger anharmonicity, the phonon lifetimes across the Brillouin zone in sLN are 1--2 orders of magnitude lower than those in cBAs, which ultimately underlies the orders-of-magnitude gap between the thermal conductivity of sLN and cBAs.

\subsection{Thermal transport in thin-film sLN}
Finally, we investigate the thermal transport properties of in thin films of sLN, motivated by their growing use in integrated photonic, acousto-optic, and electro-optic devices, where characteristic dimensions typically range from tens to hundreds of nanometers~\cite{zhu2021integrated,boes2023lithium,chen2022advances}. At these length scales, the effective thermal conductivity can deviate substantially from the bulk value because of phonon-boundary scattering~\cite{zhou2022effect}. Quantifying the intrinsic size-dependent thermal transport of sLN is therefore of both fundamental and practical importance. We evaluate this boundary-limited transport by combining the diffusive thermal conductivity $\kappa(\omega)$ obtained from HNEMD with the ballistic thermal conductance $G(\omega)$ obtained from NEMD. Their ratio defines a frequency-resolved mean free path (MFP):
\begin{equation}
\lambda(\omega) \equiv \frac{\kappa(\omega)}{G(\omega)}.
\label{eq:lambda}
\end{equation}
In the ballistic-to-diffusive crossover regime, the size dependence of thermal transport can be expressed as~\cite{dong2024molecular, Liang2025ThermalTransportaCNT}
\begin{equation}
\frac{1}{\kappa(\omega,L)} = \frac{1}{\kappa(\omega)}
\left(1 + \frac{\lambda(\omega)}{L}\right),
\label{eq:bdiff}
\end{equation}
and the thermal conductivity at a finite length $L$ is obtained by integrating over all frequencies as $\kappa(L) = \int \kappa(\omega,L) \, d\omega / 2\pi$. Here, $G(\omega)$ is evaluated from NEMD simulations at 10~K to suppress phonon-phonon scattering and realize ballistic transport conditions. We focus on room-temperature thermal
transport along the $c$-axis for various film thicknesses.

\begin{figure}[t!]
\centering
\includegraphics[width=\columnwidth]{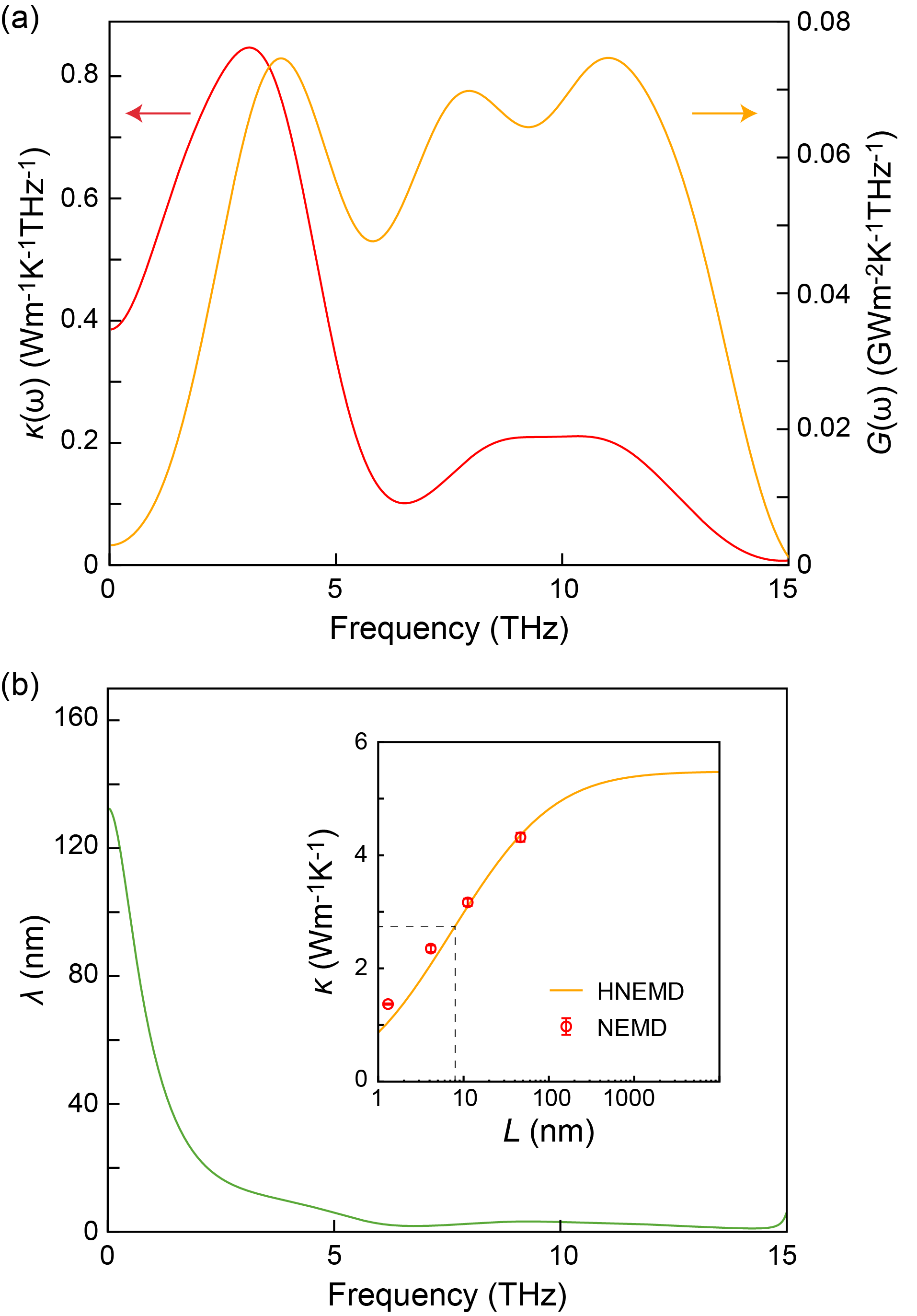}
\caption{\textbf{Phonon spectral analysis and mean free path in sLN.} (a) Spectral
  decomposition of thermal conductivity $\kappa(\omega)$ (left axis) and ballistic
  thermal conductance $G(\omega)$ (right axis). (b) Phonon mean free paths $\lambda$ as a
  function of frequency. The inset of (b) shows the size-dependent thermal conductivity
  obtained from NEMD and HNEMD simulations at different film thicknesses.
  The dashed gray lines mark the thickness at which $\kappa$ decreases to half of its bulk value.}
\label{fig:mfp}
\end{figure}

The spectral decompositions $\kappa(\omega)$ and $G(\omega)$ are shown in Fig.~\ref{fig:mfp}(a). The $\kappa(\omega)$ spectrum is dominated by acoustic phonons with frequencies below 5~THz, which serve as the primary heat carriers. In addition, a non-negligible contribution also arises from modes in the 5--15~THz spectral range, which originate from the optical phonon branches. The ballistic thermal conductance $G(\omega)$ shows substantial
  contributions across the entire 0--15~THz range, yielding a total value of $G = 1.12$~GWm$^{-2}$K$^{-1}$.

The phonon MFP spectra are calculated from Eq.~\ref{eq:lambda} using $\kappa(\omega)$ and $G(\omega)$, with the results displayed in Fig.~\ref{fig:mfp}(b). Owing to the 1--2 orders of magnitude shorter phonon lifetimes compared to cBAs, the maximum MFP in sLN reaches only approximately 140~nm. Combining this $\lambda(\omega)$ with the relation
  in Eq.~\ref{eq:bdiff}, we evaluate $\kappa(L)$ across a range of film thicknesses. For cross-validation, we also performed NEMD simulations at four different system lengths under room temperature to directly compute the thermal conductivity. These NEMD results show excellent agreement with the $\kappa(L)$ curve from Eq.~\ref{eq:bdiff}, as shown in the inset of Fig.~\ref{fig:mfp}(b) and listed in Table~\ref{tab:nemd} of the Supplemental Material~\cite{si}. This confirms that size effects become significant when the film thickness falls below 1~$\mu$m. At a thickness of 10~nm, the thermal conductivity is reduced to approximately half the bulk value. This finding provides practical guidance for thermal management in LN-based devices, where the film thickness is often comparable to or smaller than the dominant phonon MFPs.

\FloatBarrier
\section{Conclusion}

In summary, we have performed a state-of-the-art experimental and computational study of thermal transport in single-crystal stoichiometric lithium niobate. By leveraging the microscale resolution of FDTR, we effectively eliminate the influence of macroscopic structural imperfections, obtaining a room-temperature thermal conductivity of $\kappa_{ab}=4.0\pm0.8$~Wm$^{-1}$K$^{-1}$. This is in good agreement with our machine-learned MD prediction of $\kappa_{ab}=3.7\pm0.1$~Wm$^{-1}$K$^{-1}$. In addition, our MD simulations yield a room-temperature $\kappa_{c}$ of $5.6\pm0.1$~Wm$^{-1}$K$^{-1}$. Together, our robust experiments and computations definitively demonstrate the intrinsically low thermal conductivity of sLN. The simulated $\kappa$ follows a $\kappa\propto T^{-\alpha}$ scaling law with an exponent $\alpha$ near unity---a classic signature indicating the dominant role of intrinsic phonon-phonon scattering. By comparing sLN with the ultrahigh-$\kappa$ material cBAs ($\kappa\approx$ 1300~W~m$^{-1}$~K$^{-1}$) which serves as a benchmark, we find that the intrinsically low $\kappa$ of sLN does not originate from the harmonic properties: The phonon heat capacity and group velocities of sLN are comparable to or even larger than those of cBAs. Instead, due to the much larger number of atoms per primitive cell and reduced lattice symmetry, the phonon scattering phase space of sLN is expanded by 1--2 orders of magnitude compared to cBAs. Furthermore,  sLN exhibits pronounced anharmonicity, with its room-temperature anharmonic factor reaching approximately 0.27—roughly twice that of cBAs. These effects suppress phonon lifetimes by 1--2 orders of magnitude, yielding a maximum phonon mean free path of only approximately 140~nm in sLN. Consequently, size effects emerge when the film thickness falls below 1~$\mu$m, with the thermal conductivity decreasing to half the bulk value at 10~nm. Collectively, our findings establish a quantitative microscopic understanding of heat conduction in sLN and provide critical insights for thermal management of diverse lithium niobate-based devices and integrated systems.

\vspace{1em}

\begin{acknowledgments}

This work was financially supported by the Science Fund for Creative Research Groups from the National Natural Science Foundation of China (No. 52521007), the Scientific Research Innovation Capability Support Project for Young Faculty (ZYGXQNJSKYCXNLZCXM-E1) from the Ministry of Education of China, the National Key R\&D Program of China (No. 2024YFA1207900). W.Z. acknowledges support from the National Natural Science Foundation of China (No. 525B2087) and China Association for Science and Technology. Y.W. acknowledges support from the Beijing Natural Science Foundation (No. 1254054). B.S. acknowledges support  from the New Cornerstone Science Foundation through the XPLORER PRIZE. We appreciate the High-performance Computing Platform of Peking University for supporting our simulations, and acknowledge Jiangxi Unicrystal Technology Co., Ltd. for providing the LiNbO$_3$ single crystal.
\end{acknowledgments}

\bibliographystyle{apsrev4-2}

\bibliography{sLN}

\onecolumngrid
\newpage
\setcounter{figure}{0}
\renewcommand{\thefigure}{S\arabic{figure}}
\setcounter{table}{0}
\renewcommand{\thetable}{S\arabic{table}}
\begin{center}
{\large\bf Supplemental Material:}\\[8pt]
{\large\bf Intrinsically low thermal conductivity of stoichiometric lithium niobate: \\[8pt] Experimental measurement and microscopic origin}\\[12pt]
Wenjiang Zhou,\textsuperscript{1,\,$\ast$} Fuwei Yang,\textsuperscript{1,\,$\ast$} Yuxi Wang,\textsuperscript{1} Weiheng Li,\textsuperscript{1} Wujuan Yan,\textsuperscript{1} Kexin Zhang,\textsuperscript{2} and Bai Song\textsuperscript{1,\,$\dagger$}\\[8pt]
{\small
\textsuperscript{1}\textit{School of Mechanics and Engineering Science, Peking University, Beijing 100871, China}\\
\textsuperscript{2}\textit{School of Physics, Peking University, Beijing 100871, China}
}\\[8pt]
{\small $\ast$ These authors contributed equally to this work.}\\[4pt]
{\small $\dagger$ Corresponding author: songbai@pku.edu.cn}
\end{center}

\mbox{}\par
\newpage
\begin{figure}[h!]
\centering
\includegraphics{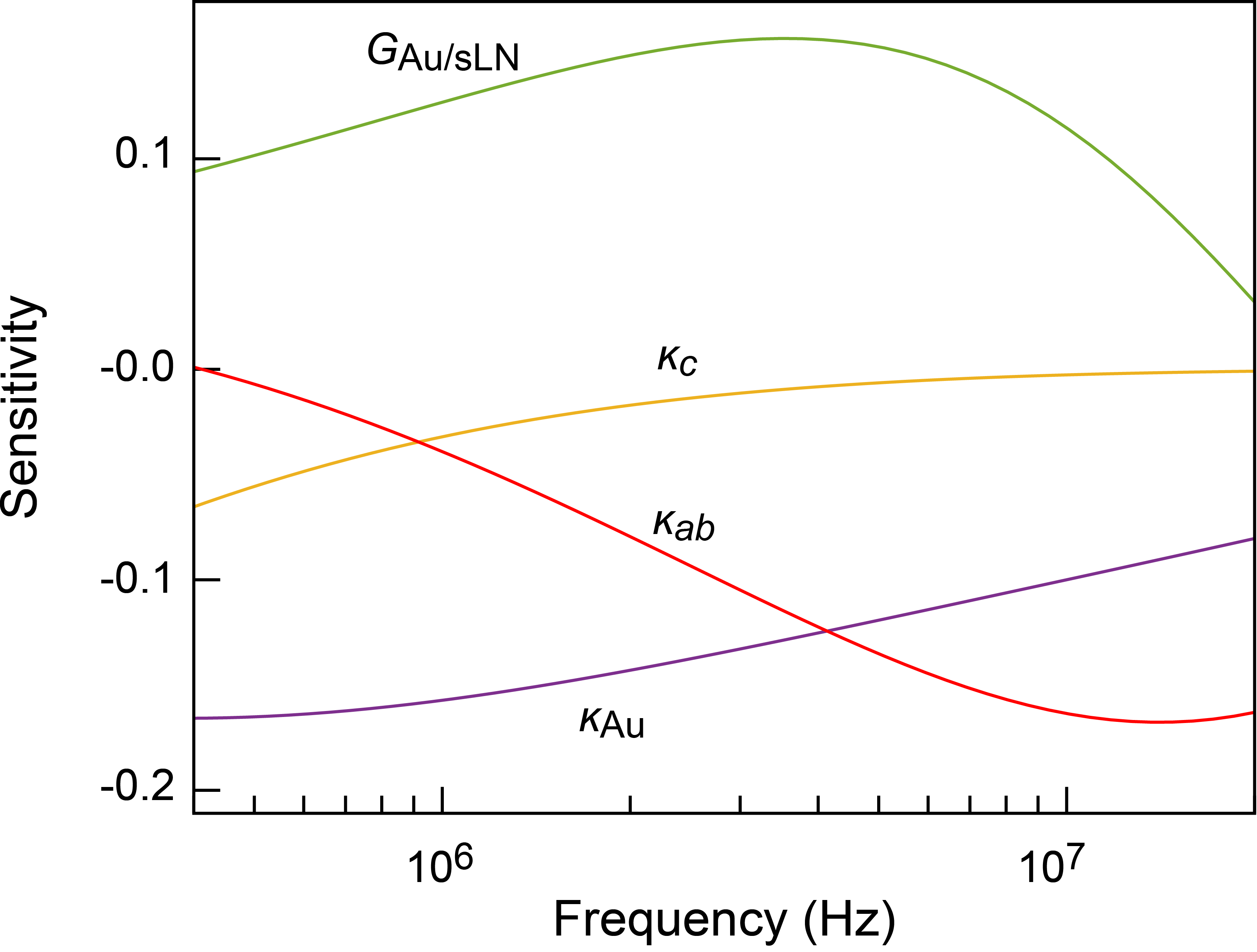}
\caption{{\bf FDTR sensitivity analysis.} The sensitivity is defined as $S_{\alpha} = \partial \ln y / \partial \ln {x_{\alpha}}$, where $y$ is the measured thermoreflectance phase and $x_{\alpha}$ is the parameter of interest. The curves show the sensitivity to the Au thermal conductivity $\kappa_{\rm Au}$, the sLN thermal conductivities $\kappa_{c}$ and $\kappa_{ab}$, and the interfacial thermal conductance $G$ between the Au transducer and the sLN sample.}
\label{fig:fdtr_sen}
\end{figure}

\newpage
\begin{figure}[h!]
\centering
\includegraphics{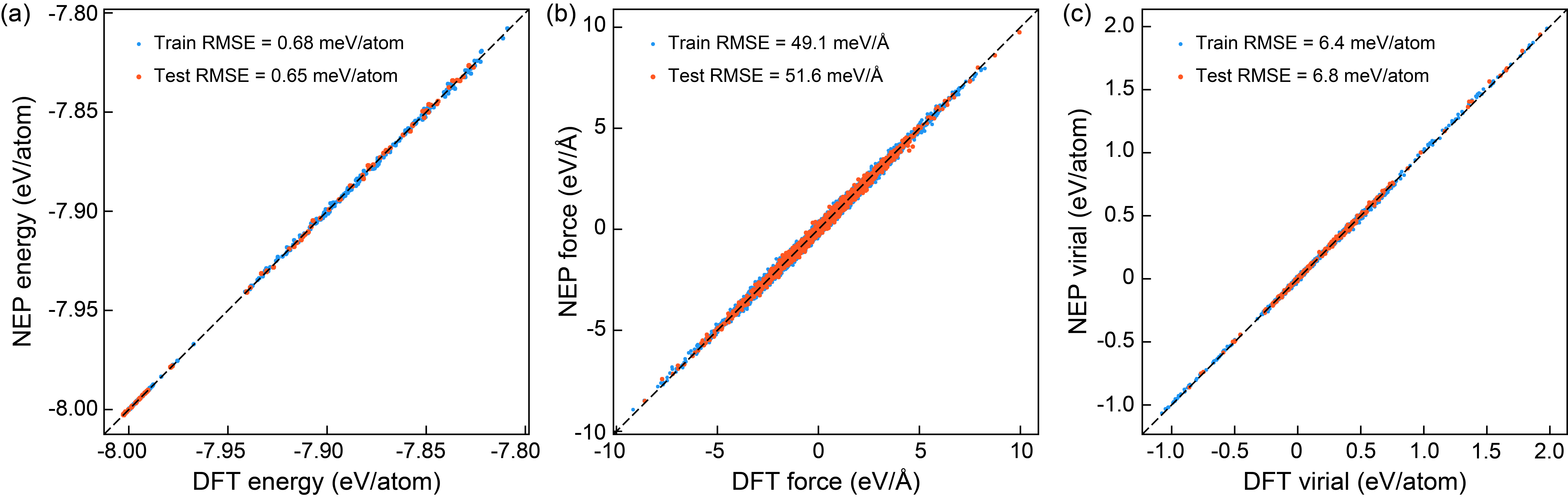}
\caption{{\bf Parity plot of the machine-learned potential.} (a) Energy, (b) force, and (c) virial components predicted by the NEP model versus the DFT reference values.}
\label{fig:parity}
\end{figure}

\newpage
\begin{figure}[h!]
\centering
\includegraphics{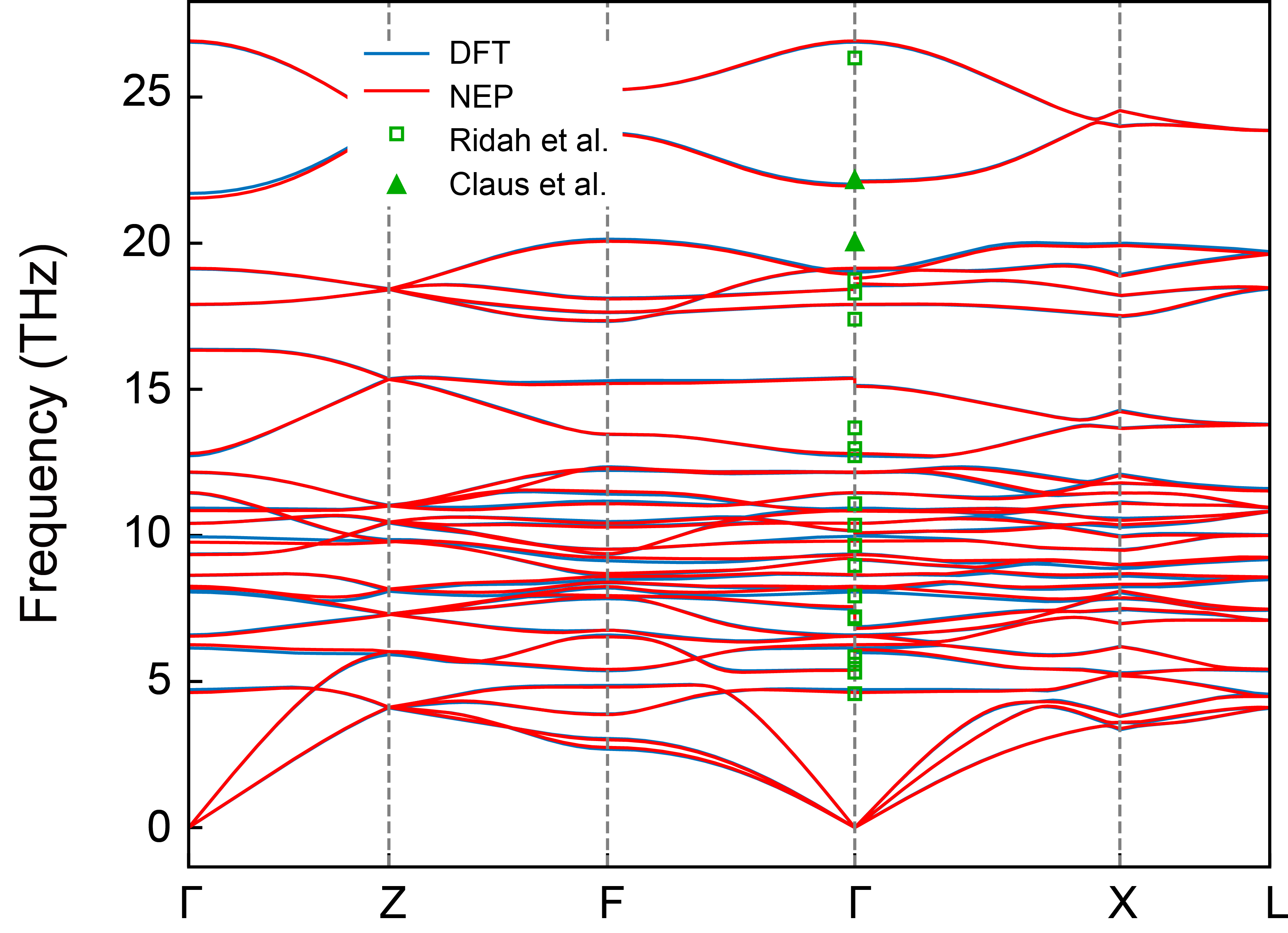}
\caption{{\bf Phonon dispersion of sLN.} The dispersion curves are calculated using the TDEP method based on DFT and NEP. Green symbols represent experimental measurements from Ridah~\textit{et~al.} and Claus~\textit{et~al.}}
\label{fig:phonon}
\end{figure}

\newpage
\begin{figure}[h!]
\centering
\includegraphics{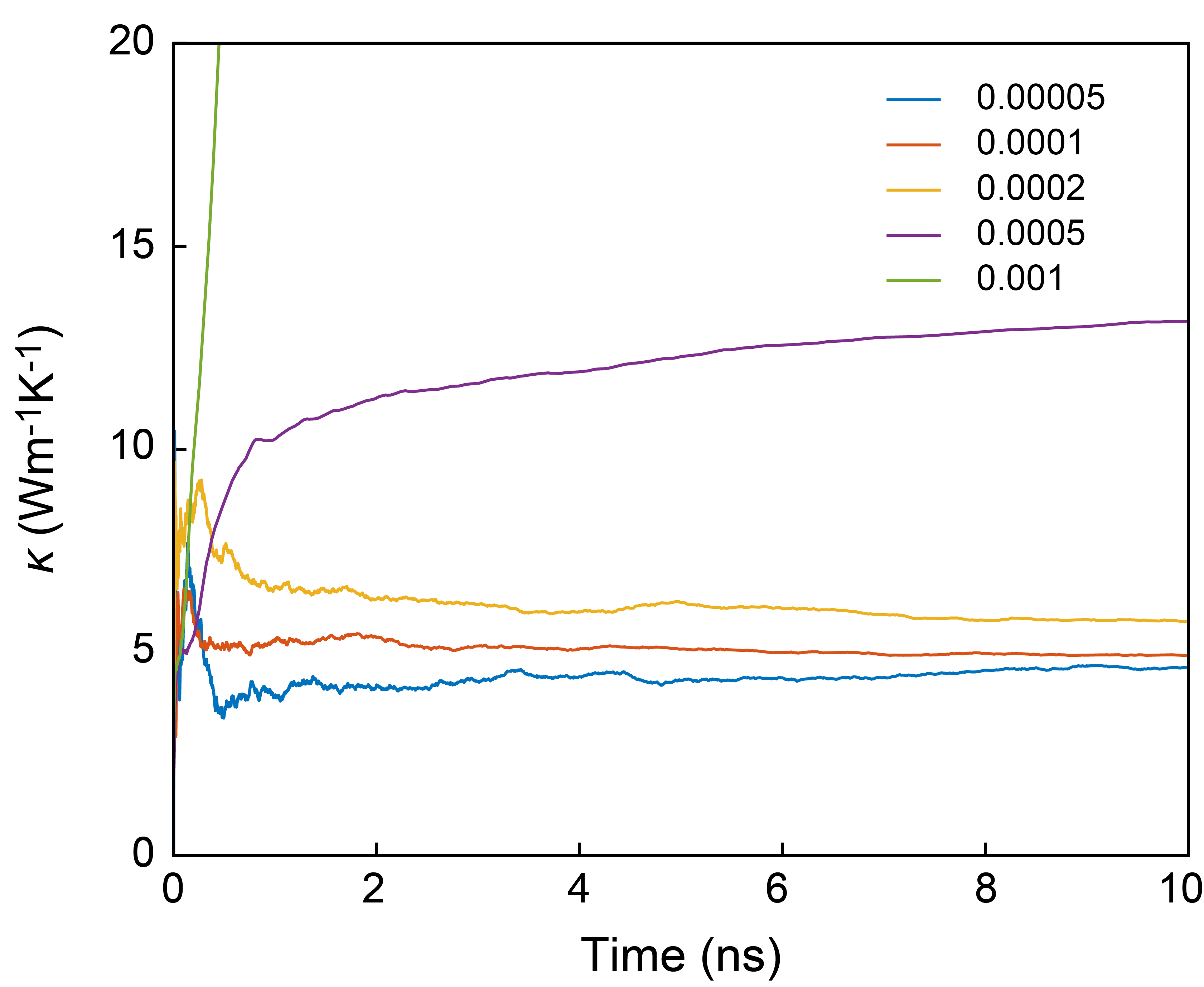}
\caption{{\bf Effect of the $\mathit{Fe}$ parameter on simulated $\kappa_{c}$ of sLN at 300~K.}}
\label{fig:hnemd_fe}
\end{figure}

\newpage
\begin{figure}[h!]
\centering
\includegraphics{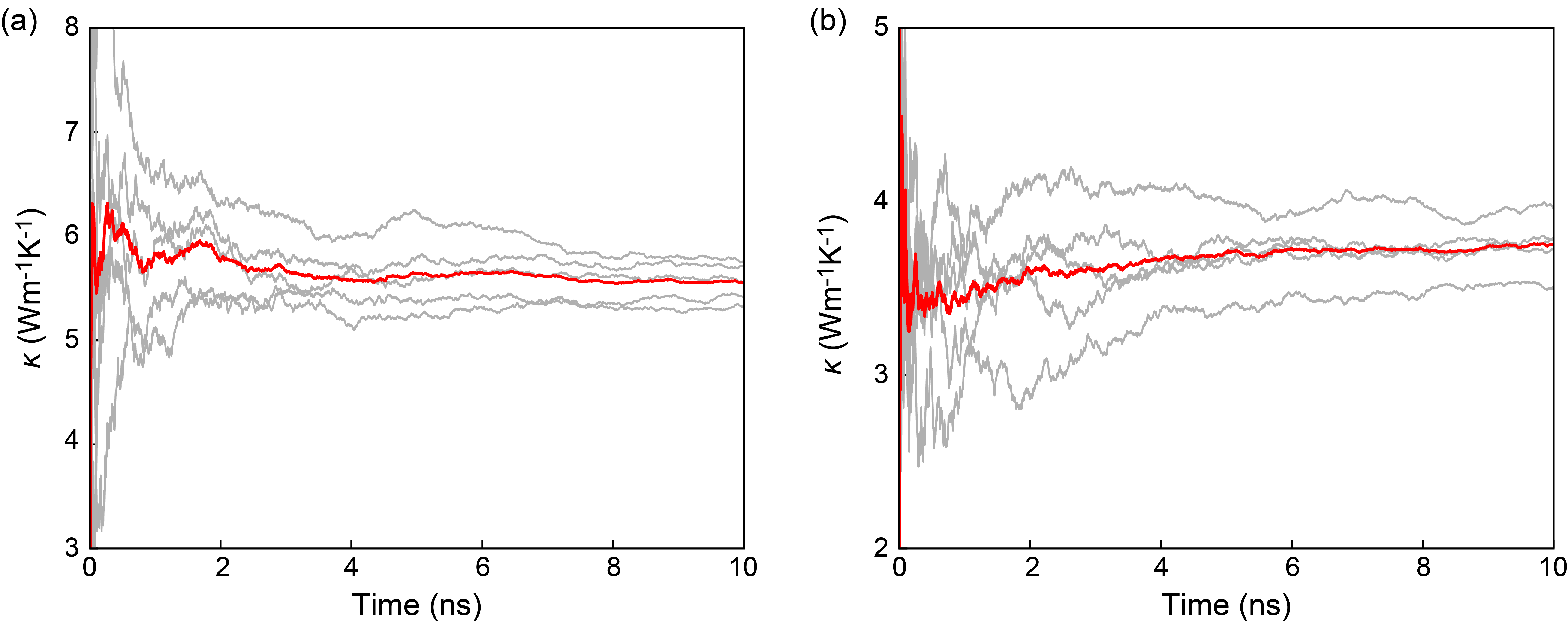}
\caption{{\bf HNEMD simulation of sLN at 300~K.} (a) Thermal conductivity $\kappa_{c}$ as a function of simulation time. (b) Thermal conductivity $\kappa_{ab}$ as a function of simulation time. Red and gray lines represent the average and individual MD simulation results, respectively.}
\label{fig:hnemd}
\end{figure}

\newpage
\begin{figure}[h!]
\centering
\includegraphics{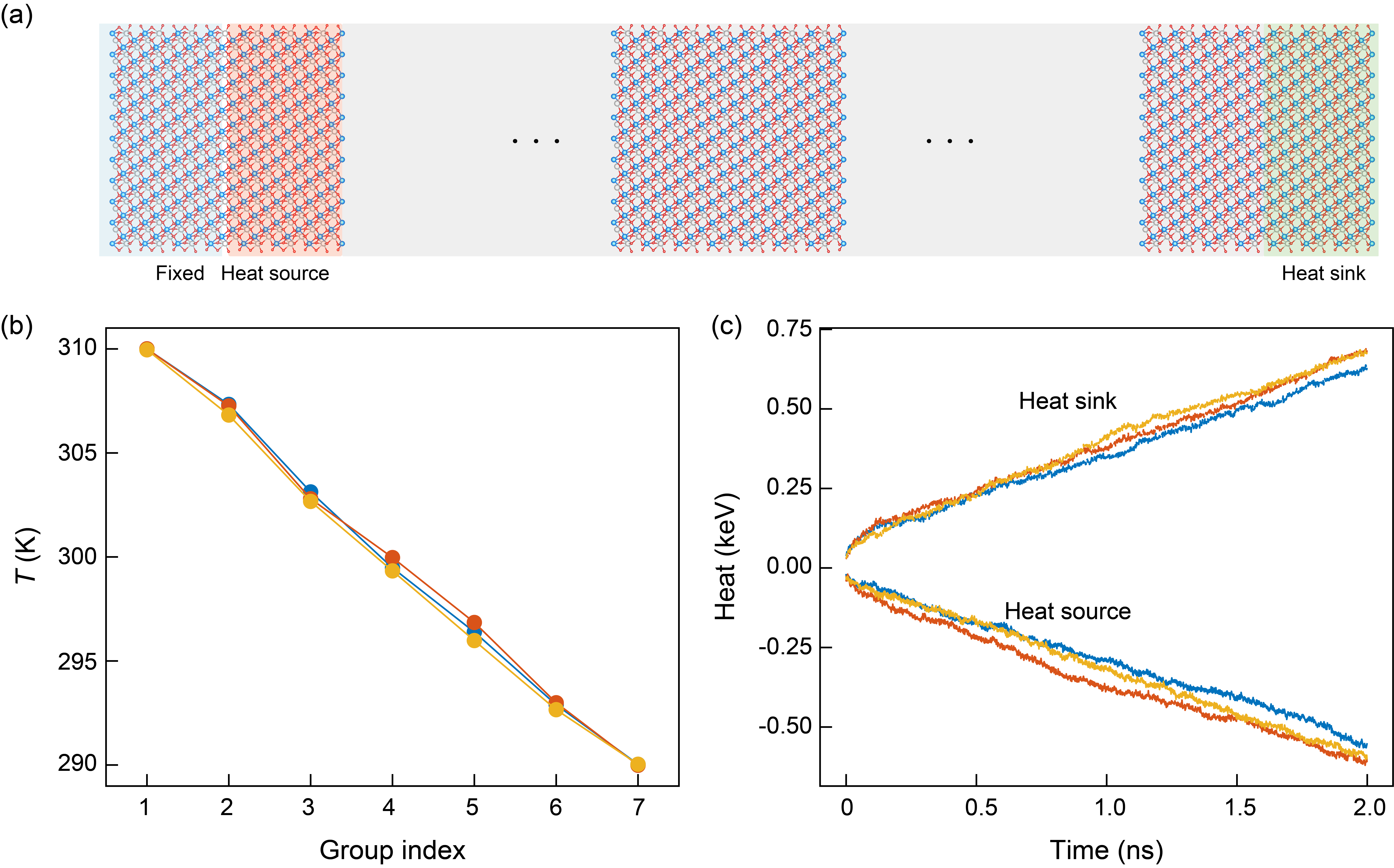}
\caption{{\bf NEMD simulation setup and analysis.} (a) Schematic of the NEMD configuration. The simulation cell is divided into four regions: fixed layers at the ends, heat source, heat sink, and the central region for temperature profile extraction and heat flux calculation. (b) Temperature profile along the $c$-axis at 300~K. Each data point represents the time-averaged temperature of a group of atoms. (c) Cumulative energy of the heat source and heat sink as a function of simulation time at 300~K. Three independent simulations are shown in both (b) and (c). The effective model length for extracting thermal conductance is $L=46.52$~nm.}
\label{fig:nemd}
\end{figure}

\newpage
{\setlength{\tabcolsep}{12pt}
\renewcommand{\arraystretch}{1.5}
\begin{table}[h]
\centering
\caption{{\bf NEMD-simulated thermal conductance of sLN at 300~K for different system lengths.} $L$ is the effective length between the heat source and heat sink. $G_1$, $G_2$, and $G_3$ represent three independent simulations.}\vspace{1em}
\label{tab:nemd}
\begin{tabular}{c c c c c}
\hline
$L$ (nm) & $G_1$ (GW\,m$^{-2}$\,K$^{-1}$) & $G_2$ (GW\,m$^{-2}$\,K$^{-1}$) & $G_3$ (GW\,m$^{-2}$\,K$^{-1}$) & $G_{\text{avg}}$ (GW\,m$^{-2}$\,K$^{-1}$)\\
\hline
1.30 & 1.042 & 1.058 & 1.052 & 1.051\\
4.13 & 0.555 & 0.583 & 0.569 & 0.569\\
11.10 & 0.276 & 0.285 & 0.287 & 0.283\\
46.52 & 0.093 & 0.091 & 0.095 & 0.093\\
\hline
\end{tabular}
\end{table}}

\newpage
{\setlength{\tabcolsep}{18pt}
\renewcommand{\arraystretch}{1.8}
\begin{table}[h]
\centering
\caption{\bf FDTR-measured thermal conductivity $\kappa_{ab}$ of sLN.}\vspace{1em}
\label{tab:fdtr}
\begin{tabular}{c c c}
\hline
$T$ (K) & $\kappa_{ab}$ (W\,m$^{-1}$\,K$^{-1}$) & Error (W\,m$^{-1}$\,K$^{-1}$)\\
\hline
183& 5.4& 1.1\\
208& 4.8& 1.0\\
238& 4.5& 0.9\\
268& 4.3& 0.9\\
298& 4.0& 0.8\\
328& 3.9& 0.8\\
358& 3.7& 0.8\\
388& 3.7& 0.8\\
\hline
\end{tabular}
\end{table}}

\newpage
{\setlength{\tabcolsep}{18pt}
\renewcommand{\arraystretch}{1.8}
\begin{table}[h]
\centering
\caption{\bf Anharmonic factor $\sigma_{\mathrm{A}}$ and thermal conductivity of representative materials at 300~K.}\vspace{1em}
\label{tab:sigma}
\begin{tabular}{c c c}
\hline
Material & $\sigma_{\mathrm{A}}$ (300~K) & $\kappa$ (W\,m$^{-1}$\,K$^{-1}$)\\
\hline
C (Diamond) & 0.09 & 3115\\
BP & 0.11 & 351\\
BN & 0.12 & 777\\
Silicon & 0.15 & 171\\
GaN & 0.15 & 210\\
BeO & 0.16 & 383\\
ZnO & 0.20 & 95\\
Ga$_2$O$_3$ & 0.21 & 9.5\\
Ga$_2$O$_3$ & 0.21 & 13.3\\
Ga$_2$O$_3$ & 0.21 & 22.5\\
InP & 0.24 & 62.1\\
PbTe & 0.44 & 2.20\\
cBAs & 0.13 & 1300\\
LiNbO$_3$ ($\kappa_{c}$) & 0.27 & 5.6\\
LiNbO$_3$ ($\kappa_{ab}$) & 0.27 & 3.7\\
\hline
\end{tabular}
\end{table}}

\clearpage
\twocolumngrid

\end{document}